\documentclass[a4paper,useAMS,usenatbib,11pt]{article}
\usepackage[margin=2.5cm]{geometry}
\usepackage{enumitem}
\usepackage{graphicx}                    
\usepackage{natbib}                     
\usepackage{amssymb}                    
\usepackage{amsmath}
\usepackage{color}                       
\usepackage{url}                         

\newcommand{\vct}[1]{\ensuremath\boldsymbol{#1}}

\begin{document}


\title{Power Anisotropy in the Magnetic Field Power Spectral Tensor of Solar Wind Turbulence}


\author{R. T. Wicks$^{1}$\thanks{E-mail: r.wicks@imperial.ac.uk} , M. A. Forman$^{2}$, T. S. Horbury$^{1}$, S. Oughton$^{3}$\\$^{1}$ Space and Atmospheric Physics Group, Imperial College London, SW7 2AZ, UK.\\$^{2}$ Stony Brook University, Stony Brook, NY11790-3800 USA. \\$^{3}$ Department of Mathematics, University of Waikato, Hamilton, New Zealand}


\maketitle
\label{firstpage}

\begin{abstract}We observe the anisotropy of the power spectral tensor of magnetic field fluctuations in the fast solar wind for the first time. In heliocentric {\it RTN} coordinates the power in each element of the tensor has a unique dependence on the angle between the magnetic field and velocity of the solar wind ($\theta_B$) and the angle of the vector in the plane perpendicular to the velocity ($\phi_B$). We derive the geometrical effect of the high speed flow of the solar wind past the spacecraft on the power spectrum in the frame of the plasma $P(\vct{k})$ to arrive at the observed power spectrum $P(f,\theta_B,\phi_B)$ based on a scalar field description of turbulence theory. This allows us to predict the variation in the $\phi_B$ direction and compare it to the data. We then transform the observations from {\it RTN} coordinates to magnetic-field-aligned coordinates. The observed reduced power spectral tensor matches the theoretical predictions we derive in both {\it RTN} and field-aligned coordinates which means that the local magnetic field we calculate with wavelet envelope functions is an accurate representation of the physical axis of symmetry for the turbulence and implies that on average the turbulence is axi-symmetric. We also show that we can separate the dominant toroidal component of the turbulence from the smaller but significant poloidal component and that these have different power anisotropy. We also conclude that the magnetic helicity is anisotropic and mostly two-dimensional, arising from wavevectors largely confined to the plane perpendicular to $\vct{B}$.
\end{abstract}

%

\section{Introduction}

Fast solar wind from the poles of the Sun is an excellent example of MHD turbulence, with the fluctuations being approximately incompressible \citep{Goldstein95, Horbury05}. The Ulysses spacecraft provides a unique data set with extended periods in this continuous fast polar solar wind \citep{Ebert09} and high cadence magnetic field data \citep{Balogh}. Such observations allow us to investigate how turbulence makes the nominally collisionless solar wind behave like a gas with shocks and structures, and why superthermal particles and cosmic rays appear to be diffusively coupled to the solar wind, allowing exchange of energy. The details of that coupling are not yet understood completely, and the poorly understood anisotropy of the turbulence is a part of the problem. Recently measurements have been made using the Ulysses data clearly showing the importance of the magnetic field direction in the turbulence \citep{Horbury08, Podesta09, Luo10, Wicks10}, with different power amplitudes and spectral indices in different directions relative to the local mean magnetic field. Attempts have been made to choose between theories of anisotropic turbulent cascades \citep[e.g.][]{GS95, GS97, Boldyrev06, Lithwick07} by observing the scaling of the power at different angles to the mean field. 
\par
All of these studies have concentrated on the trace of the magnetic power spectral tensor (i.e. the total power in magnetic fluctuations) rather than the whole tensor, which is needed to fully describe turbulence. The second order correlation tensor and the associated power spectral tensor are central parts of generalized turbulence theories \citep{Robertson, Batchelor46, BatchelorBook, Chandrasekhar50}. Incompressible MHD turbulence is different in many respects to incompressible hydrodynamic turbulence, primarily since it has two solenoidal fields, $\vct{V}$ and $\vct{B}$ \citep{Chandrasekhar51a, Chandrasekhar51b, Biskamp03}. Theoretical treatments show that solenoidal fields in MHD plasmas (e.g. $\nabla \cdot \vct{B} = 0$) require correlation and power spectral tensors which are completely described by four standard tensor forms multiplying four scalar functions \citep{Oughton97}. 
\par
Frequency power spectra from single-spacecraft observations $P(f)$ are equivalent to the `reduced' form of the full three-dimensional wave vector power spectrum $P(\vct{k})$  \citep{FredricksCoroniti76, Forman11}. In this context, $\vct{k}$ represents the wavelength and orientation of the 3D spatial structure of the turbulence, which is advected past the spacecraft at supersonic speeds \citep{Taylor}. The resulting integral is a type of tomographic projection called a Radon transform \citep{Radon, Debnatha}. It is impossible to separate wave vectors that have the same projection on the direction of flow \citep{FredricksCoroniti76, Forman11}. This causes a permanent ambiguity in the observed power spectrum and means that {\it in-situ} observations by single spacecraft can only fully resolve the 3D spectrum or the related correlation tensor if they are isotropic. Multiple-spacecraft missions, such as Cluster, have been used to overcome this problem \citep[e.g.][]{Osman07, Narita10, Sahraoui10} but do not spend long in the solar wind and so are difficult to use for turbulence studies which require ensemble averages over large data sets, and the resolution of wave vectors is relatively coarse. 
\par
In Section 2 we use the Ulysses data to make the first measurements of all nine elements of the reduced magnetic power spectral tensor $P_{ij}(f, \hat{\vct{b}})$ as a function of the direction $\hat{\vct{b}}$ of the local mean magnetic field in the solar wind. These are measured in the Sun-spacecraft aligned {\it RTN} coordinates \citep[see][]{Burlaga84, Franz02} and show a remarkable amount of variation with $\hat{\vct{b}}$. In Section 3 we derive the dependence of the reduced $P_{ij}(f, \hat{\vct{b}})$ on $P_{ij}(\vct{k})$ using the general tensor formalism of \cite{Oughton97} and show how the resulting 4 scalar functions appear in the measured power spectral tensor in {\it RTN} coordinates. This may seem inelegant but the observations we wish to understand are made in this coordinate system. In Sections 4 and 5 we use the observed variation of power with $\hat{\vct{b}}$ in {\it RTN} coordinates to show that on average the turbulent fluctuations are axi-symmetric and elliptically polarized. The polarization ellipse of the ensemble average of turbulent fluctuations is aligned along unit vector axes we define which are themselves aligned with respect to $\vct{B}$.
\par
Finally we convert the observed power spectral tensor from {\it RTN} into magnetic-field-aligned coordinates and compare to the derived reduced power spectral tensor in this coordinate system. Combining the results in both coordinate systems allows us to demonstrate that the locally averaged magnetic field is an accurate representation of the axis of symmetry of the turbulence and therefore to plot the true reduced power anisotropy of the magnetic field. We show that the poloidal scalar function, which includes all pseudo-Alfv\'{e}nic fluctuations, can be separated from the toroidal function, which includes all shear Alfv\'{e}nic fluctuations, in field-aligned coordinates and that they have different magnitudes and power anisotropy. We also show that the magnetic helicity is predominantly in fluctuations with wave vectors near to the plane perpendicular to $\vct{B}$. These properties represent important tests that any turbulence theory must satisfy and the results presented here are important for all kinds of astrophysical turbulence: the solar wind, solar dynamo, and interstellar, galactic and intergalactic magnetic fields. 
\begin{table*}[tb]
	\centering
		\begin{tabular}{ccccccc}
		Distance (AU)&Latitude ($^{\circ}$)&$\left|\vct{B}\right|$ (nT)&$\left|\vct{V}\right|$ (km/s)&$\rho$ (cm$^{-3}$)&V$_A$ (km/s)&$\beta_i$\\
		$2$ - $2.28$&$79$ - $74.8$&$1.5 \pm 0.3$&$780 \pm 20$&$0.5 \pm 0.1$&$45 \pm 8$& $1.6 \pm 0.6$ \\
		\end{tabular}
		\caption{Spacecraft location and average solar wind conditions for the 50 days of Ulysses data used in the analysis.}
		\label{table:solarwind}
\end{table*}

\section{Measuring the Reduced Power Tensor}

We use one-second resolution magnetic field data from the Ulysses spacecraft from days 200 to 249 (inclusive) of 1995 when the spacecraft was in a continuous polar fast stream characteristic of high latitudes at solar minimum. The location of the spacecraft and average solar wind conditions for this period are summarised in Table \ref{table:solarwind}. 
\par
We measure all components of the power spectral tensor of magnetic fluctuations using a complex Morlet wavelet decomposition of the time series reported in {\it RTN} coordinates \citep{Horbury08, Podesta09, Wicks10}. Wavelet coefficients $w_i$ are calculated using the inverse Fourier transform of the Fourier representation of the Morlet wavelet \citep{TorrenceCompo, Podesta09} with the Fourier transform of the magnetic field $\tilde{B_i}(\omega)$,
\begin{equation}
w_{i}(s(f), t) = \left(\frac{2\pi s}{\delta t}\right)^{1/2} \int_{-\infty}^{\infty}{\tilde{B_i}(\omega) \, \pi^{-1/4} H(\omega) \, e^{-(s\omega - \omega_0)^2/2} \, e^{2\pi i t \omega} \, \mathrm{d}\omega}
\label{eq:WaveletTransform}
\end{equation}
with $\delta t$ the time cadence of the data (1 second), $s$ is the wavelet (time) scale which is varied to select different frequencies $f$, related by $s = \frac{\omega_0 + \sqrt{2 + \omega_0^2}}{4\pi f}$, $\omega_0 = 6$, $H(\omega)$ is the Heaviside step function, and $i$,$j$ run over R,T,N. The measured power spectral tensor as a function of frequency is then:
\begin{equation}
P_{ij}(f, t) = w_i(f, t)w_j^*(f, t)
\label{eq:WaveletPower}
\end{equation}
Note that the wavelet amplitudes $w_i$ contain phase information and are complex, making $P_{ij}$ a Hermitian tensor. Each power measurement $P_{ij}(f, t)$, can be associated with the direction of the mean magnetic field $\vct{B}(t)$ calculated using the same averaging envelope at time $t$. The field direction is defined by the angles $\theta_B$ and $\phi_B$ as shown in Figure \ref{fig:1}: 
\begin{equation}
\vct{B}
     =
        |\vct{B}|\cos\theta_B \hat{\vct{R}}
      + |\vct{B}|\sin\theta_B \cos\phi_B \hat{\vct{T}}
      + |\vct{B}|\sin\theta_B \sin\phi_B \hat{\vct{N}} \label{eq:thetaphidef}
\end{equation}
and we define:
\begin{equation}
\hat{\vct{b}} = \frac{\vct{B}}{|\vct{B}|}
\end{equation}
\begin{figure*}[tb]%
\centering
\includegraphics[width=\textwidth]{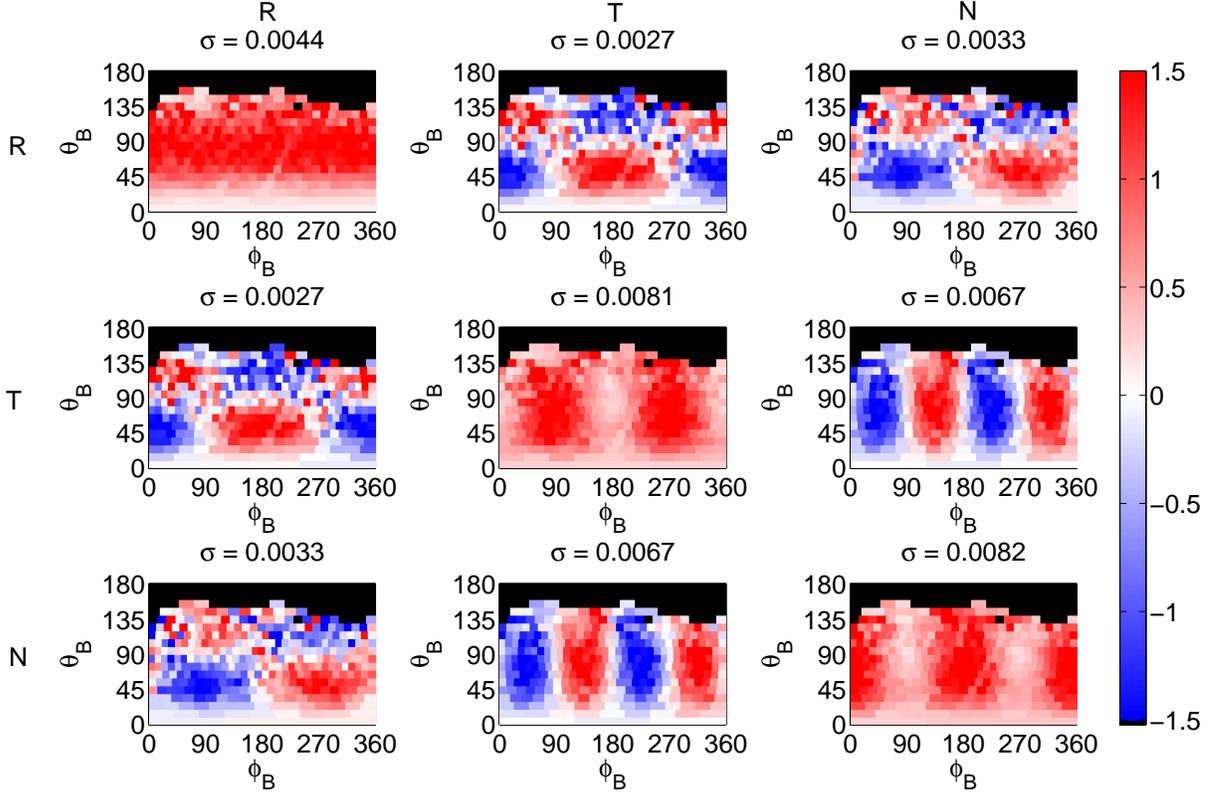}
\caption{Real component of the power spectral tensor from Ulysses magnetic field data at $f = 0.098$ Hz. Black areas represent bins that have fewer than 10 points in them. Red represents positive and blue negative contributions to the power, with white being zero. The colour scale has been scaled to the standard deviation of all points contributing to the power in each map individually, the value of which is shown above each panel.}%
\label{fig:3}%
\end{figure*}
\par
Although {\it RTN} coordinates are more awkward theoretically than field-aligned coordinates for understanding magnetic turbulence, we use {\it RTN} to simplify data handling. The power contributions are accumulated and averaged in 404 separate direction bins: 18 equally wide 10 degree bins in $\theta_B$, and variable width bins in $\phi_B$ to keep the solid angle area of each bin approximately constant. We keep bins equally spaced in $\theta_B$ since we are interested in the behaviour in this direction for physical reasons, thus where $\theta_B$ is near $ 0^{\circ}$ or $180^{\circ}$ there are fewer bins in $\phi_B$. The mean and standard error of each of the mean power contributions $P_{ij} = w_iw_j^*$ in each bin are then associated with the $\theta_B$ and $\phi_B$ at the center of the bin. Thus $P_{ij}(f, t)$ is converted into $P_{ij}(f, \hat{\vct{b}})$ by this averaging process.
\par
\begin{figure*}[htb]%
\centering
\includegraphics[width=\textwidth]{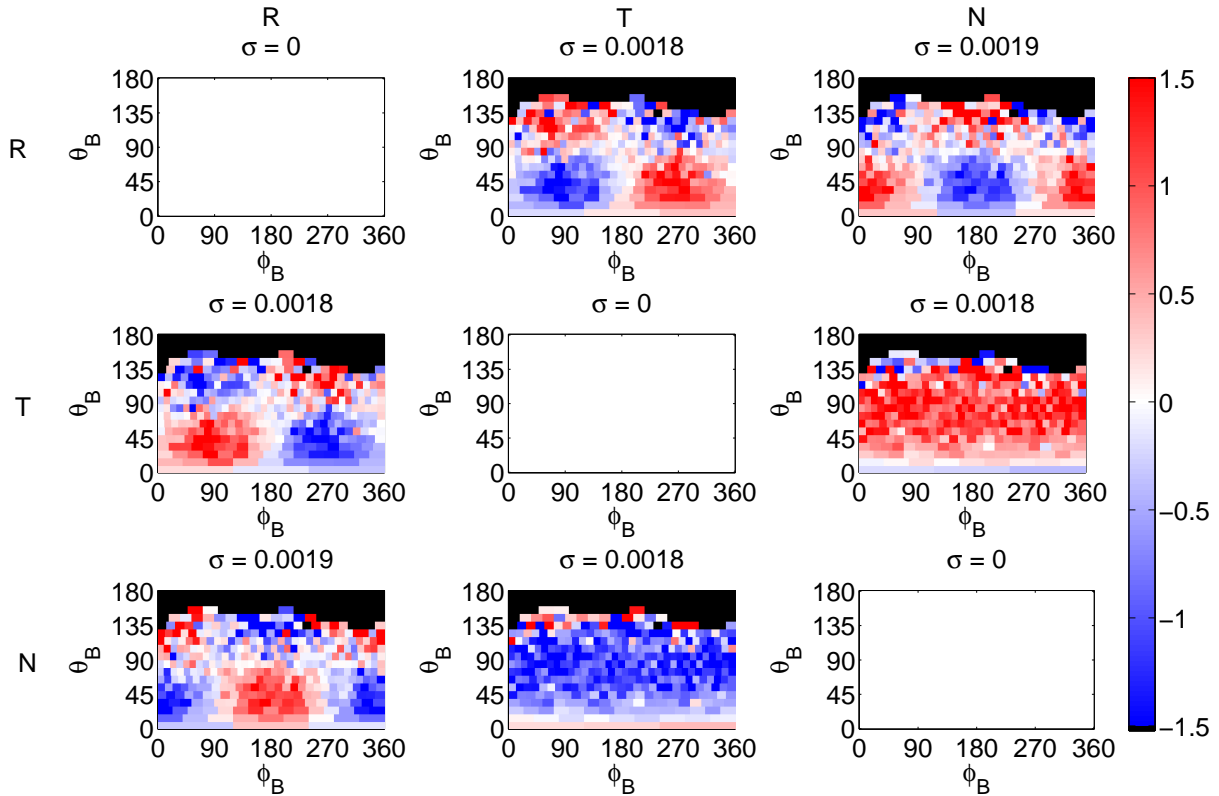}
\caption{Imaginary component of the power spectral tensor from Ulysses magnetic field data at $f = 0.098$ Hz. Black areas represent bins that have fewer than 10 points in them. Red represents positive and blue negative contributions to the power, with white being zero. The colour scale has been scaled to the standard deviation of all points contributing to the power in each map individually, the value of which is shown above each panel. Note that the diagonal components are zero by construction.}%
\label{fig:4}%
\end{figure*}

Using this method a map can be made of the power distributed over $\theta_B$ and $\phi_B$ for each of the 9 tensor elements of Equation \ref{eq:WaveletPower} at each wavelet scale. We have measured the anisotropy of the power spectral tensor at a range of frequencies ($0.25\times10^{-2} < f < 0.25$ Hz), which allows us to verify the results presented here as typical over the inertial range of turbulence. In this paper we concentrate on a single scale since we are interested in power anisotropy; we will return to the scaling with $f$ of the power spectral tensor in a future publication. Examples of the distribution of power in real and imaginary parts for each tensor element are shown in Figures \ref{fig:3} and \ref{fig:4}, at a frequency of $f = 0.098$ Hz, which is at the high frequency end of the anisotropic inertial range, maximizing the observable power anisotropy, as shown in \cite{Wicks10}. 
\par
\begin{figure*}[htb]%
\centering
\includegraphics[width=\textwidth]{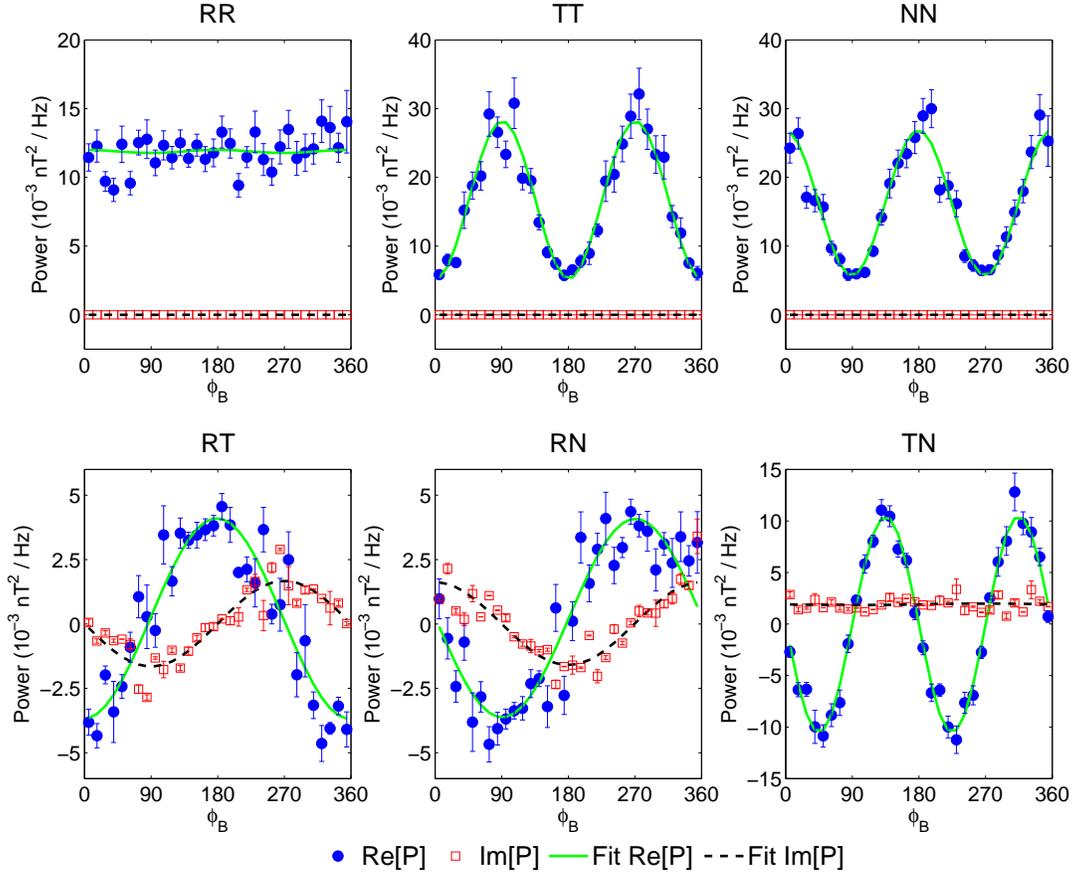}
\caption{The $\phi_B$ dependence of power at $\theta_B = 65^{\circ}$ and $f = 0.098$ Hz. The blue and red points are the real and imaginary data respectively and the green and black lines are their sinusoidal fits as described in Equations \ref{eq:fitsstart} - \ref{eq:fitsend} with the values quoted in Table \ref{table:fitvalues}.}%
\label{fig:Phi_Fits}%
\end{figure*}

By definition the diagonal terms of the power tensor are real, since they are the wavelet coefficient multiplied by its complex conjugate. The off-diagonal terms are complex and have both real and imaginary parts. The data are presented in Figures \ref{fig:3} and \ref{fig:4} as a 2D map of the surface of a sphere, the horizontal direction in each of the nine plots is the $\phi_B$ direction and has variable bin-width, and the vertical direction is the $\theta_B$ direction and has fixed bin width of 10 degrees. The color scale runs from dark blue for the largest negative values of power, through white at $P = 0$ and then to dark red for the largest positive values of power. The colour scale is shown at the side of the plots and is calculated individually for each map in terms of the standard deviation $\sigma$ of all data contributing to that map (this includes the systematic and sinusoidal variations and so is considerably larger than the standard deviation in any individual bin, as shown in Figure \ref{fig:Phi_Fits}). The colour can be scaled onto the absolute value of the power by the value of $\sigma$ in nT$^2$/Hz shown above each map. The off-diagonal maps appear noisier than those on the diagonal because their generally smaller magnitude makes the errors proportionally larger and they have both positive and negative regions with zero in between making any uncertainty in these regions appear more clearly in the color map. There are also fewer points per bin on average for $\theta_B > 90^{\circ}$ making the error and therefore the scatter proportionally larger in this region.
\par
We will return to the absolute values of the data later, but for now we note that the maps have different, but clear, harmonic variations with $\phi_B$. Within experimental error it appears that the RT and RN components are first harmonics of $\phi_B$, their amplitudes are equal and they are $90^{\circ}$ out of phase. The TT, NN and the real part of the TN components are second harmonics of $\phi_B$ with similar amplitudes and multiples of $45^{\circ}$ out of phase with each other. In Figure \ref{fig:4} the imaginary part of the TN element varies in only the $\theta_B$ direction. Another striking feature of Figure \ref{fig:4} is that the standard deviation of each map, used to calibrate the colour scale, is almost equal across all maps. The tensor is Hermitian by construction which provides the mirror symmetry about the diagonal. In fact, the dramatic dependence on $\phi_B$ is an artifact of using {\it RTN} coordinates, although it can be modified by the presence of non-axi-symmetric turbulence. In Section 3 we show how this arises, how it can be removed and how it can be used to extract extra information about the structure of the turbulence. 
\par
We can quantify the $\phi_B$ dependence of the tensor components by fitting functions to the observed variation of power with $\phi_B$. We fit sinusoidal functions of $\phi_B$ to each tensor element containing a contribution from the $\hat{\vct{R}}$ direction and sinusoidal functions of $2\phi_B$ to the others, at each $f$ and $\theta_B$, with a non-linear least squares fitting method to determine the fitting parameters at each $f$ and $\theta_B$ as specified in Equations \ref{eq:fitsstart} - \ref{eq:fitsend}. The resemblance of these sinusoidal functions to the data provides motivation for employing them in fits. Their suitability is deeper than this, however, as we prove in Sections 3 and 4. The fits all consist of a real constant average $A$ independent of $\phi_B$ and a real sinusoidal amplitude $B$ as well as an imaginary constant average $C$ and an imaginary sinusoidal amplitude $D$; each sinusoidal function also has a phase offset, $E$ for the real part and $F$ for the imaginary. This process is repeated for all values of $\theta_B$ at which there are five or more points to fit to. As $\theta_B$ coverage is limited by the reduction in solid angle close to $\theta_B = 0$, this means that the angle range covered is $15^{\circ} < \theta_B < 175^{\circ}$, at each frequency, so there are never more fitting parameters (3) than data points (a minimum of 5), although the fits with $\theta_B$ closest to zero are the least accurate. Thus each of the six independent elements of the tensor at each $f$ and most $\theta_B$ can be described by these scalar parameters with separate averages, amplitudes and phase shifts for the real and imaginary parts.
\par 
\begin{align}
P_{RR}(\phi_B) =& A_{RR} + B_{RR}\sin(\phi_B + E_{RR}) \label{eq:fitsstart}\\
P_{TT}(\phi_B) =& A_{TT} + B_{TT}\cos(2\phi_B + E_{TT}) \\
P_{NN}(\phi_B) =& A_{NN} + B_{NN}\cos(2\phi_B + E_{NN}) \\
P_{RT}(\phi_B) =& A_{RT} + B_{RT}\cos(\phi_B + E_{RT}) + i(C_{RT} + D_{RT}\sin(\phi_B + F_{RT})) \\
P_{RN}(\phi_B) =& A_{RN} + B_{RN}\sin(\phi_B + E_{RN}) + i(C_{RN} + D_{RN}\cos(\phi_B + F_{RN})) \\
P_{TN}(\phi_B) =& A_{TN} + B_{TN}\sin(2\phi_B + E_{TN}) + i(C_{TN} + D_{TN}\sin(2\phi_B + F_{TN})) \label{eq:fitsend}
\end{align}
\par
\begin{table*}[htb]
	\centering
		\begin{tabular}{ccccccc}
			Tensor element & A & B & C & D \\
			$RR$ & $13.0 \pm 0.5$ & $-0.1 \pm 0.8$ &  &  \\
			$TT$ & $15.8 \pm 0.8$ & $-11 \pm 1$ &  &  \\
			$NN$ & $16.5 \pm 0.7$ & $11.6 \pm 0.9$ &  &  \\
			$RT$ & $0.1 \pm 0.4$ & $-2.4 \pm 0.5$ & $0.03 \pm 0.2$ & $-1.0 \pm 0.3$ \\
			$RN$ & $0.3 \pm 0.4$ & $-2.7 \pm 0.6$ & $-0.05 \pm 0.2$ & $1.1 \pm 0.3$ \\
			$TN$ & $0.09 \pm 0.4$ & $-11.0 \pm 0.6$ & $2.2 \pm 0.2$ & $-0.1 \pm 0.3$ \\
		\end{tabular}
		\caption{Values for power in units of $10^{-3}$ nT$^2$Hz$^{-1}$ of the fitted parameters in Equations \ref{eq:fitsstart} - \ref{eq:fitsend} at $\theta_B = 65^{\circ}$ and $f = 0.098$ Hz, the same data as in Figure \ref{fig:Phi_Fits}. The fitted angular phase is not shown since all are within errors of 0. Note that within errors there are only 4 independent real values and two independent imaginary values.}
		\label{table:fitvalues}
\end{table*}
Figure \ref{fig:Phi_Fits} shows a typical example of how the real and imaginary parts of each element of the power spectral tensor at a certain $\theta_B$, vary with $\phi_B$. We can see that $\mathit{Re}[P_{RT}]$ and $\mathit{Re}[P_{RN}]$ elements behave to a very close approximation like $\cos\phi_B$ and $\sin\phi_B$ respectively implying that $E \sim 0$. Similarly the real part of $P_{TN}$, and $P_{NN}$ and $P_{TT}$ behave to a very close approximation like $\sin(2\phi_B)$ and $\cos(2\phi_B)$ respectively, again implying that $E \sim 0$. We quantify this further by looking at the measured phase shifts $E$ from the fitting of the real sinusoidal functions. The average shift of the phase in $\phi_B$ over all scales and angles is only $\bar{E} = 0.6 \pm 1.1 ^{\circ}$ and always in the range $\pm10^{\circ}$. This is smaller than the angular resolution of the method we use $(10^{\circ})$ and so within the accuracy of the method we cannot distinguish $E$ from $0$.
\par
Looking at Figure \ref{fig:Phi_Fits} again, we see that the imaginary part of the power varies to a very good approximation like $\pm\sin\phi_B$ in the $P_{RT}$ and $P_{TR}$ elements and $\pm\cos\phi_B$ in $\mathit{Im}[P_{RN}]$ and $\mathit{Im}[P_{NR}]$. This implies that $F \sim 0$ and again we quantify this by looking at the phase shifts $F$ of the imaginary parts of $P_{RT}$ and $P_{RN}$, which are too small to measure using our technique, being $F = 0.2 \pm 1.0^{\circ}$ and always in the range $\pm10^{\circ}$. This average ignores the $\mathit{Im}[P_{TN}]$ component since there is no sinusoidal variation and so these values of $F$ are poorly constrained.
\par
Table \ref{table:fitvalues} shows the fitted parameters for the data in Figure \ref{fig:Phi_Fits} at $f = 0.098$ Hz and $\theta_B = 65^{\circ}$. There are only 4 measurably distinct non-zero real parameters ($A_{RR}$, $A_{TT} = A_{NN}$, $B_{TT} = -B_{NN} = B_{TN}$, and $B_{RT} = B_{RN}$) and two non-zero imaginary parameters ($D_{RT} = -D_{RN}$ and $C_{TN}$). This is true at all $f$ and $\theta_B$ in the range we studied. This means that there are at most six functions of $\theta_B$ at each $f$ which together completely describe the properties of the power spectral tensor of the turbulence. We show in the next section that this organization follows from geometry, the solenoidal field, and the conversion from power spectra in wave vector $\vct{k}$ to power spectra of the time series as seen at the spacecraft by deriving six $\phi_B$-independent functions corresponding to the six $\phi_B$-independent values observed in the data. 

\section{Systematic Effects of Geometry}

\begin{figure}[htb]%
\centering
\includegraphics[width=5cm]{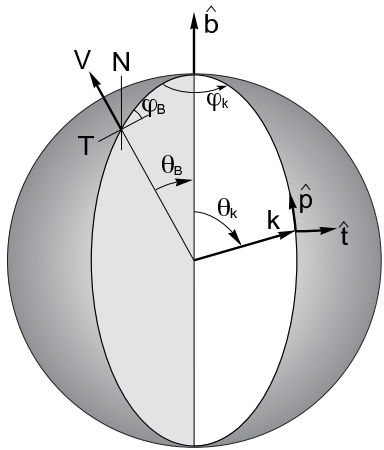}
\caption{The geometry used in this paper, $\vct{V}$ is the solar wind velocity, considered to be in the radial (R) direction, $\hat{\vct{b}}$ is the unit vector of the magnetic field, $\hat{\vct{k}}$ is the unit wave vector of a fluctuation. R, T and N are heliocentric coordinates, $\hat{\vct{t}}$ and $\hat{\vct{p}}$ are toroidal and poloidal directions, $\theta_B$ and $\phi_B$ are angular coordinates of $\hat{\vct{b}}$,  and $\theta_k$ and $\phi_k$ are the angles between $\hat{\vct{k}}$ and the $(\vct{V},\hat{\vct{b}})$ plane.}
\label{fig:1}%
\end{figure}

There are several potential problems in comparing observations with theory in solar wind turbulence studies. One is that all {\it in-situ} spacecraft observations of the solar wind are of a `reduced' spectrum, but theory usually addresses the spectrum in $\vct{k}$ space. Power spectra calculated from time series of single point observations made in a fast flowing medium carrying a relatively slowly evolving turbulence are, by Taylor's hypothesis \citep{Taylor}, an integral of the $P_{ij}(\vct{k})$ in wave vector space over the plane perpendicular to the flow defined by $\vct{k}\cdot\vct{V} = 2\pi f$ \citep{FredricksCoroniti76}:
\begin{equation}
P_{ij}(f, \vct{V}) = \iiint \! P_{ij}(\vct{k}) \delta(2\pi f - \vct{k} \cdot \vct{V}) \, \mathrm{d}^{3}\vct{k}.
\label{eq:ReducedPower}
\end{equation}
When $P_{ij}(\vct{k})$ is anisotropic, $P_{ij}(f, \vct{V})$ will depend on the direction of $\vct{V}$ relative to any symmetry in $P_{ij}(\vct{k})$. If symmetry in $P_{ij}(\vct{k})$ is organized by the direction of the local mean magnetic field, $\hat{\vct{b}}$, $P_{ij}(f, \vct{V})$ can be better written as $P_{ij}(f, \hat{\vct{b}})$. In fact the total (trace) power $P(f, \hat{\vct{b}}) = \sum{P_{ii}(f, \hat{\vct{b}})}_{i = RTN}$ is known to be anisotropic in both power and spectral index as a function of $\theta_B$ \citep{Horbury08, Podesta09, Luo10, Wicks10, Forman11}. In order to derive the expected geometrical effect of the reduction on the spectrum we must account for three vectors $\left( \vct{k}, \vct{B}, \vct{V} \right)$ and their corresponding coordinate systems aligned with $\hat{\vct{b}}$ and $\vct{V}$. Figure \ref{fig:1} shows these vectors, $\vct{V}$ the solar wind velocity, $\hat{\vct{b}}$ the unit vector of the magnetic field, both of which are measured in {\it RTN} coordinates in this analysis, and $\hat{\vct{k}} $ the unit wave vector of a turbulent fluctuation. 
\par
The magnetic field is solenoidal, $\nabla \cdot \vct{B} = 0$, ensuring that all fluctuations are confined to the plane perpendicular to $\vct{k}$, which is tangent to the surface of the sphere at position $\vct{k}$ in Figure \ref{fig:1}. To describe the fluctuations of $\delta \vct{B}$ in this plane we define the toroidal direction $\hat{\vct{t}}$, perpendicular to both $\vct{k}$ and $\hat{\vct{b}}$, and the poloidal direction $\hat{\vct{p}}$, perpendicular to $\vct{k}$ and $\hat{\vct{t}}$ thus:
\begin{eqnarray}
\hat{\vct{t}} & = & \frac{ \hat{\vct{b}} \times \hat{\vct{k}} }{ \left| \hat{\vct{b}} \times \hat{\vct{k}} \right| } \label{eq:t} \\
\hat{\vct{p}} & = & \hat{\vct{k}} \times \hat{\vct{t}} \label{eq:p}
\end{eqnarray}
Considering the sphere in polar coordinates with $\hat{\vct{b}}$ as
the polar axis and $\vct{k}$ as the radius vector:
$\hat{\vct{p}}$ is in the direction of decreasing $\theta_k$, and
$\hat{\vct{t}}$ is in the direction of increasing $\phi_k$;
$(\hat{\vct{k}}, \hat{\vct{t}}, \hat{\vct{p}})$ is a
right-handed coordinate system.
\par
The power spectral tensor of toroidal fluctuations alone is a scalar function $\mathit{Tor}(\vct{k})$ times the dyadic $\left[\hat{\vct{t}}:\hat{\vct{t}}\right]$, and for poloidal fluctuations alone is $\left[\hat{\vct{p}}:\hat{\vct{p}}\right] \mathit{Pol}(\vct{k})$.  If both polarizations exist, any correlation between them will result in the additional power spectral elements $\left[\hat{\vct{t}}:\hat{\vct{p}} + \hat{\vct{p}}:\hat{\vct{t}}\right]\mathit{C}(\vct{k})$ which is real and $\mathit{i}\left[\hat{\vct{t}}:\hat{\vct{p}} - \hat{\vct{p}}:\hat{\vct{t}}\right]k\mathit{H}(\vct{k})$ which is imaginary and anti-symmetric. The four scalar functions are all real and the spectral tensor structure corresponds in detail to the complete description of transverse light waves with the Stokes parameters \citep{Chandrasekhar60} ($I = \mathit{Tor} + \mathit{Pol}, Q = \mathit{Tor} - \mathit{Pol}, U = 2\mathit{C}, V = 2k\mathit{H}$), and to the description of solenoidal MHD fluctuations by \cite{Oughton97} where our $\mathit{Tor}(\vct{k})$, $\mathit{Pol}(\vct{k})$, $\mathit{C}(\vct{k})$ and $\mathit{H}(\vct{k})$ correspond to $E$, $E - (\hat{\vct{b}} \times \vct{k})^2 F$, $k(\hat{\vct{b}} \times \vct{k})^2\mathit{C}$ and $\mathit{H}$ in their paper. The toroidal fluctuations are perpendicular to $\hat{\vct{b}}$, and so they are sometimes called `Alfv\'{e}nic', since this is the polarization of small-amplitude shear Alfv\'en waves. Similarly, the $\mathit{Pol}(\vct{k})$ fluctuations are sometimes called `pseudo-Alfv\'{e}nic' because their polarization is the same as that of small amplitude pseudo-Alfv\'en waves \citep[c.f.][]{ChoLazarian02}. Following these theoretical structures we define the anisotropic power tensor as a function of $\vct{k}$:
\begin{equation}
P(\vct{k}) = \mathit{Tor}(\vct{k})\left[\hat{\vct{t}}:\hat{\vct{t}}\right] + \mathit{Pol}(\vct{k})\left[\hat{\vct{p}}:\hat{\vct{p}}\right] + \mathit{C}(\vct{k})\left[\hat{\vct{t}}:\hat{\vct{p}} + \hat{\vct{p}}:\hat{\vct{t}}\right] + i k\mathit{H}(\vct{k})\left[ \hat{\vct{t}}:\hat{\vct{p}} - \hat{\vct{p}}:\hat{\vct{t}} \right].
\label{eq:PowerFrom4Scalars}
\end{equation}
This formalism is completely general and Equation \ref{eq:PowerFrom4Scalars} describes any turbulent field satisfying the solenoidal condition regardless of any symmetry.
\par
In order to use Equations (\ref{eq:ReducedPower} - \ref{eq:PowerFrom4Scalars}) with the measured tensor in {\it RTN} coordinates we must express $\hat{\vct{t}}$ and $\hat{\vct{p}}$ in {\it RTN} coordinates, but keep $\vct{k}$ in field-aligned coordinates. For this we define a new coordinate system aligned with $\hat{\vct{b}}$ and containing the radial flow direction of the solar wind:
\begin{eqnarray}
\vct{e}_z & = & \hat{\vct{b}} \label{eq:ez} \\
\vct{e}_y & = & \frac{\vct{e}_z \times \vct{V}}{\left|\vct{e}_z \times \vct{V}\right|} = \frac{\vct{e}_z \times \vct{R}}{\sin \theta_B} \label{eq:ey} \\
\vct{e}_x & = & \vct{e}_y \times \vct{e}_z \label{eq:es} \\
\vct{k} & = & k_x\vct{e}_x + k_y\vct{e}_y + k_z\vct{e}_z
\label{eq:kses}
\end{eqnarray}
Defining $\vct{e}_y$ this way means $\vct{V}$ is in the $x$-$z$ plane and thus $\vct{V}\cdot\vct{k} = |\vct{V}| \left(\sin\theta_B k_x + \cos\theta_B k_z \right)$. We will exploit the fact that $k_y$ is therefore not in the delta function of Equation \ref{eq:ReducedPower} to help us understand the symmetries of the scalar functions later. It should also be noted that $\phi_B$ is not in the delta function either, thus $\phi_B$ can potentially be moved outside the integral in Equation \ref{eq:ReducedPower}.
\par
We find $\hat{\vct{t}}$ and $\hat{\vct{p}}$ in {\it RTN} coordinates for any $\vct{k}$ using Equations \ref{eq:ez} - \ref{eq:kses} in Equations \ref{eq:t} and \ref{eq:p} with $k_{\perp} = \sqrt{k_x^2 + k_y^2}$ and $k = \sqrt{k_x^2 + k_y^2 + k_z^2}$:
\begin{equation}
\hat{\vct{t}} = \left( \begin{array}{c}
-\frac{k_y}{k_{\perp}} \sin\theta_B \hat{\vct{R}}\\
\left(\frac{k_x}{k_{\perp}} \sin\phi_B + \frac{k_y}{k_{\perp}} \cos\theta_B\cos\phi_B\right) \hat{\vct{T}}\\
\left(-\frac{k_x}{k_{\perp}} \cos\phi_B + \frac{k_y}{k_{\perp}} \cos\theta_B\sin\phi_B\right) \hat{\vct{N}}
\end{array} \right),
\label{eq:tarray}
\end{equation}
\begin{equation}
\hat{\vct{p}} = \left( \begin{array}{c}
\left(\frac{k_{\perp}}{k}\cos\theta_B - \frac{k_xk_z}{kk_{\perp}}\sin\theta_B \right) \hat{\vct{R}}\\
\left(\frac{k_{\perp}}{k}\sin\theta_B\cos\phi_B - \frac{k_z}{kk_{\perp}}\left(k_y\sin\phi_B - k_x\cos\theta_B\cos\phi_B \right) \right) \hat{\vct{T}}\\
\left(\frac{k_{\perp}}{k}\sin\theta_B\sin\phi_B + \frac{k_z}{kk_{\perp}}\left(k_y\cos\phi_B + k_x\cos\theta_B\sin\phi_B \right) \right) \hat{\vct{N}}
\end{array} \right)
\label{eq:parray}
\end{equation}
Note that $\phi_B$ does not appear in the {\bf R} component of either $\hat{\vct{t}}$ or $\hat{\vct{p}}$, and only as $\sin\phi_B$ or $\cos\phi_B$ in the other two components. Since $\phi_B$ is not involved in the integration in Equation \ref{eq:ReducedPower} this dependence appears directly in the maps of $P_{ij}(f, \hat{\vct{b}})$ as 0, $1^{st}$, or $2^{nd}$ order harmonics of $\phi_B$, and is easily seen in the maps of $P_{ij}(f, \hat{\vct{b}})$ in Figures \ref{fig:3} and \ref{fig:4}. The amplitude of each of these harmonics in $\phi_B$ is an integral involving $\theta_B$ over the power distribution $P(\vct{k})$ and is a function of $\theta_B$. We will now determine what these amplitude functions are by gathering terms with no dependence, first harmonic and second harmonic dependence on $\phi_B$.

\section{Harmonics of $\phi_B$}

There are only six independent linear combinations of the elements of $P^{RTN}(f, \hat{\vct{b}})$ which have no, first or second harmonic $\phi_B$ dependence. This is exactly the same number as the independent power amplitudes of the fitted sinusoidal functions of $\phi_B$ observed in Table \ref{table:fitvalues}. These can be expressed in terms of the four scalar functions by putting Equation \ref{eq:PowerFrom4Scalars} through the reduction integral (Equation \ref{eq:ReducedPower}) with the $\hat{\vct{t}}$ and $\hat{\vct{p}}$ vectors in {\it RTN} coordinates, as in Equations \ref{eq:tarray} and \ref{eq:parray}. The results are projections of the four scalar functions from $\vct{k}$-space onto the {\it RTN} coordinate system which we now derive.
\par
The Trace and $P_{RR}$ are independent of $\phi_B$ and Equation \ref{eq:ReducedPower} gives the two projection integrals:
\begin{align}
P_{RR}(f, \theta_B) = & & & \nonumber\\
 \iiint \! \Bigl( t_R^2&\mathit{Tor}(\vct{k}) + p_R^2\mathit{Pol}(\vct{k}) + 2t_Rp_R\mathit{C}(\vct{k}) \Bigr) \delta(2\pi f - \vct{k} \cdot \vct{V}) \, \mathrm{d}^{3}\vct{k}, & & \label{eq:PRRNoPhi} \\
 &&&\nonumber\\
\text{Trace}(f, \theta_B) =  & \sum_{i=RTN}{P_{ii}(f, \hat{\vct{b}})} & & \nonumber\\
 = & \iiint \! \Bigl( \mathit{Tor}(\vct{k}) +\mathit{Pol}(\vct{k}) \Bigr) \delta(2\pi f - \vct{k} \cdot \vct{V}) \, \mathrm{d}^{3}\vct{k}, & & \label{eq:TraceNoPhi}
\end{align}
We then collect the real components which are first harmonics of $\phi_B$, if we combine them we find:
\begin{align}
I_1 \  e^{\mathit{i} \phi_B} = & \  \textit{Re}\left(P_{RT}(f, \hat{\vct{b}})\right) + \mathit{i} \textit{Re}\left(P_{RN}(f, \hat{\vct{b}})\right)  = \nonumber\\
e^{\mathit{i}\phi_B} \, \iiint \! \Bigl( t_R& Z_t \mathit{Tor}(\vct{k}) +  p_R Z_p \mathit{Pol}(\vct{k}) + (t_R Z_p + p_R Z_t)\mathit{C}(\vct{k}) \Bigr) \delta(2\pi f - \vct{k} \cdot \vct{V}) \, \mathrm{d}^{3}\vct{k} \label{eq:PRTRNNoPhi}
\end{align}
where we have expressed the $\phi_B$ dependence of $\hat{\vct{t}}$ and $\hat{\vct{p}}$ compactly in the following way:
\begin{eqnarray}
Z_t(\theta_B, \vct{k})e^{\mathit{i}\phi_B} = & t_T + \mathit{i}t_N = & e^{\mathit{i}\phi_B}\frac{1}{k_{\perp}}\left(k_y \cos\theta_B - \mathit{i}k_x \right) \label{eq:Zt} \\
Z_p(\theta_B, \vct{k})e^{\mathit{i}\phi_B} = & p_T + \mathit{i}p_N = & e^{\mathit{i}\phi_B}\frac{1}{kk_{\perp}}\left(k_{\perp}^2 \sin\theta_B + k_xk_z \cos\theta_B + \mathit{i}k_yk_z \right) \label{eq:Zp}
\end{eqnarray}
Note that $Z_t$, $Z_p$ are complex functions of $\theta_B$ and $\vct{k}$ so $I_1$ is complex and the absolute phase of $P_{RT}$ and $P_{RN}$ are important clues to finding $\mathit{Tor}(\vct{k})$, $\mathit{Pol}(\vct{k})$ and $\mathit{C}$. 
\par
Similarly collecting the real components which are second harmonic in $\phi_B$ we find:
\begin{eqnarray}
I_2 \ e^{2\mathit{i}\phi_B} = P_{TT}(f, \hat{\vct{b}}) - P_{NN}(f, \hat{\vct{b}}) + \mathit{i}\left(P_{TN}(f, \hat{\vct{b}}) + P_{NT}(f, \hat{\vct{b}})\right) = \nonumber\\ e^{2\mathit{i}\phi_B} \, \iiint \! \Bigl( Z_t^2 \mathit{Tor}(\vct{k}) + Z_p^2 \mathit{Pol}(\vct{k}) + 2Z_pZ_t\mathit{C}(\vct{k}) \Bigr) \delta(2\pi f - \vct{k} \cdot \vct{V}) \, \mathrm{d}^{3}\vct{k}.
\label{eq:PTTNNTNNoPhi}
\end{eqnarray}
this integral is also complex and so the absolute phase is also important.
\par
From the imaginary part of $P^{RTN}(f, \hat{\vct{b}})$ we find there are only two equations with simple dependence on $\phi_B$, both projections of the scalar function $\mathit{H}(\vct{k})$. The tensor form multiplying $\mathit{H}(\vct{k})$ in Equation \ref{eq:PowerFrom4Scalars} is:
\begin{equation}
\hat{\vct{t}}_i\hat{\vct{p}}_j - \hat{\vct{p}}_i\hat{\vct{t}}_j = \epsilon_{ijm}(\hat{\vct{t}} \times \hat{\vct{p}})_m = \epsilon_{ijm}\hat{\vct{k}}_m
\end{equation}
so that
\begin{equation}
\mathit{Im}\left[P_{ij}(f, \hat{\vct{b}})\right] = \epsilon_{ijm} \iiint \! \vct{k}_m\mathit{H}(\vct{k}) \delta(2\pi f - \vct{k}\cdot\vct{V}) \mathrm{d}^{3}\vct{k}.
\end{equation}
Combining the two off-diagonal terms with first harmonic dependence on $\phi_B$:
\begin{align}
I_3 \ e^{\mathit{i} \phi_B} & = \mathit{Im}[P_{RT}(f, \hat{\vct{b}})] + \mathit{i}\mathit{Im}[P_{RN}(f, \hat{\vct{b}})] \nonumber \\
 & = e^{\mathit{i}\phi_B} \iiint \! \left(t_RZ_p - p_RZ_t\right)k\mathit{H}(\vct{k}) \delta(2\pi f - \vct{k} \cdot \vct{V}) \mathrm{d}^{3}\vct{k} & \label{eq:I3}
\end{align}
this integral is potentially useful as it can give us information about the symmetries of $\mathit{H}(\vct{k})$ through the different projections of $\hat{\vct{t}}$ and $\hat{\vct{p}}$ it contains.
\par
Finally, we find that one imaginary term has no dependence on $\phi_B$:
\begin{equation}
\mathit{Im}\left[P_{TN}(f, \hat{\vct{b}})\right] = \iiint \! (\vct{k}\cdot\vct{R})\mathit{H}(\vct{k}) \delta(2\pi f - \vct{k}\cdot\vct{V}) \mathrm{d}^{3}\vct{k},
\end{equation}
and by noticing that $\vct{k}\cdot\vct{V} = |V|(\vct{k}\cdot\vct{R})$ we easily recover the well known and frequently used result of \cite{Matthaeus82} that $\mathit{Im}[P_{TN}]$ is the reduced magnetic helicity:
\begin{equation}
\frac{H_m}{2} = \mathit{Im}\left[P_{TN}(f, \hat{\vct{b}})\right] = \frac{2\pi f}{V}\iiint \! \mathit{H}(\vct{k}) \delta(2\pi f - \vct{k}\cdot\vct{V}) \mathrm{d}^{3}\vct{k}.
\label{eq:Hm}
\end{equation}
\par
We have not yet made any assumptions about the symmetry or behaviour of the four scalar functions and so these six relations are generally true. We now look at the Ulysses data in more detail to see what restrictions the observed tensor elements place on the four scalar functions.

\section{Observational restrictions on the scalar fields}

We can now return to the observations in Figures \ref{fig:3}-\ref{fig:Phi_Fits} and Table \ref{table:fitvalues} and compare them to the geometrical effects derived above. We can rearrange Equations \ref{eq:PRTRNNoPhi} and \ref{eq:PTTNNTNNoPhi} to illustrate the dependence of the individual tensor elements on $\phi_B$ and the integrals $I_1$ and $I_2$ which are functions of $\theta_B$ and $f$. We drop the dependence on $f$ and $\theta_B$ temporarily for simplicity:
\begin{eqnarray}
\textit{Re}[P_{RT}] & = &\textit{Re}[I_1]\cos(\phi_B) - \textit{Im}[I_1]\sin(\phi_B) \label{eq:RT}\\
\textit{Re}[P_{RN}] & = &\textit{Re}[I_1]\sin(\phi_B) + \textit{Im}[I_1]\cos(\phi_B) \label{eq:RN}\\
\textit{Re}[P_{NT}] & = &\frac{1}{2}\left(\textit{Re}[I_2]\sin(2\phi_B) + \textit{Im}[I_2]\cos(2\phi_B)\right) \label{eq:TN}\\
\textit{Re}[P_{TT} - P_{NN}] & = & \textit{Re}[I_2]\cos(2\phi_B) - \textit{Im}[I_2]\sin(2\phi_B) \label{eq:TTNN}
\end{eqnarray} 
A similar rearrangement of the imaginary part using Equation \ref{eq:I3} yields:
\begin{eqnarray}
\textit{Im}[P_{RT}] & = &\textit{Re}[I_3]\cos(\phi_B) - \textit{Im}[I_3]\sin(\phi_B) \label{eq:ImRT}\\
\textit{Im}[P_{RN}] & = &\textit{Re}[I_3]\sin(\phi_B) + \textit{Im}[I_3]\cos(\phi_B) \label{eq:ImRN}
\end{eqnarray}
\par
A surprising property of the data described earlier is that the angular phase offsets $E \sim 0$ and $F \sim 0$ for all of the measured quantities. By comparing Equations \ref{eq:fitsstart} - \ref{eq:fitsend} with $E$ set to $0$ with Equations \ref{eq:RT} and \ref{eq:RN} we can see that in the solar wind this implies that $\textit{Im}[I_1] \sim 0$ and in addition using Equations \ref{eq:TN} and \ref{eq:TTNN} $\textit{Im}[I_2] \sim 0$. $F \sim 0$ and Equations \ref{eq:ImRT} and \ref{eq:ImRN} similarly imply that $\textit{Re}[I_3] \sim 0$. The implications of this are discussed later in this section.
\begin{figure*}[htb]%
\centering
\includegraphics[width=\textwidth]{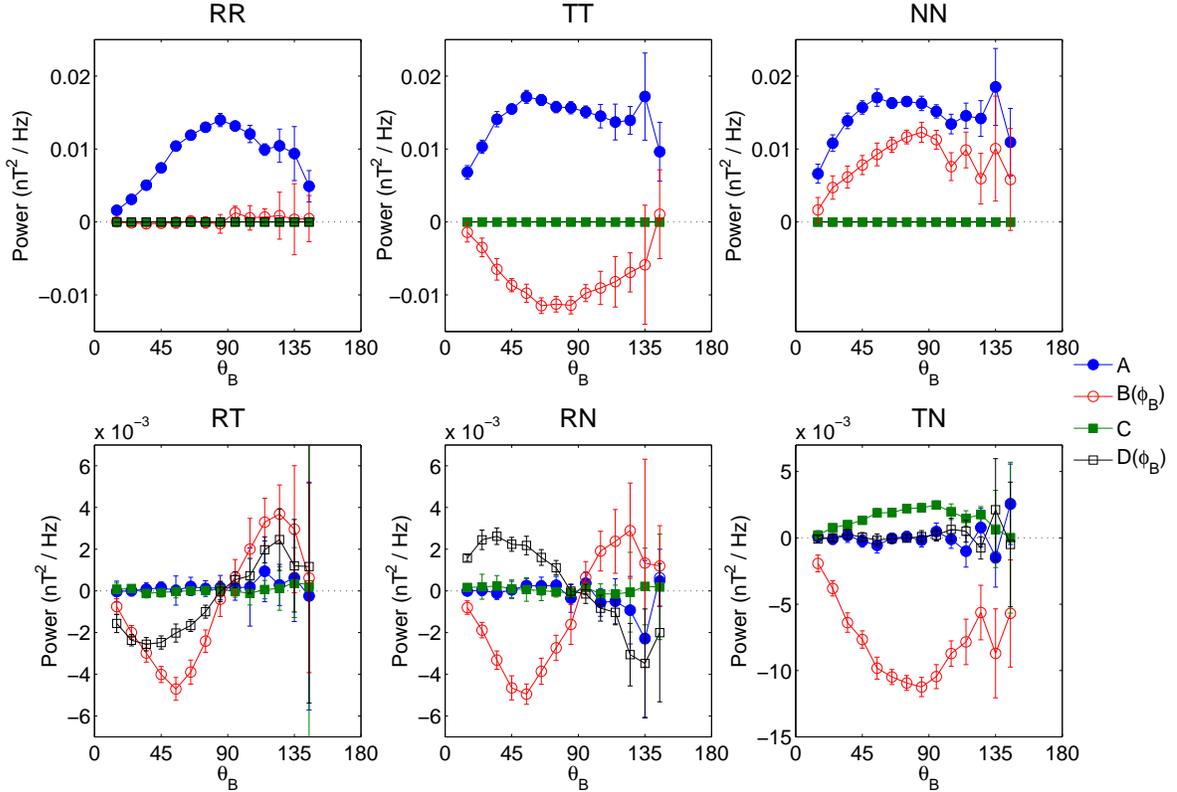}
\caption{The amplitude of the fitting parameters $A$, $B$, $C$, and $D$ for the six independent power spectral tensor elements as a function of $\theta_B$, at $f = 0.098$ Hz.}%
\label{fig:Theta_Residuals}%
\end{figure*}
\par
Equations \ref{eq:startobservedPs} - \ref{eq:endobservedPs} show the resulting $\theta_B$ and $\phi_B$ dependencies of each element of the power tensor measured in {\it RTN} coordinates on the six $\phi_B$-independent integrals defined in the previous section. These functions fit the observations very well as can be seen by comparing the equations to Figures \ref{fig:3}, \ref{fig:4} and \ref{fig:Phi_Fits}. Just as in Table \ref{table:fitvalues} there are only four real and two imaginary amplitudes. These terms are a result of the geometry shown in Figure \ref{fig:1} and the reduced nature of the measurements as described by Equation \ref{eq:ReducedPower} as well as the turbulent power spectrum. 
\begin{align}
			P_{RR} = & P_{RR}(\theta_B) \label{eq:startobservedPs}\\
			P_{TT} = & \frac{1}{2}\Bigl(\text{Trace}(\theta_B) - P_{RR}(\theta_B) + I_2(\theta_B)\cos(2\phi_B)\Bigr) \\
			P_{NN} = & \frac{1}{2}\Bigl(\text{Trace}(\theta_B) - P_{RR}(\theta_B) - I_2(\theta_B)\cos(2\phi_B)\Bigr) \\
			P_{RT} = & I_1(\theta_B)\cos(\phi_B) - iI_3(\theta_B)\sin(\phi_B) \\
			P_{RN} = & I_1(\theta_B)\sin(\phi_B) + iI_3(\theta_B)\cos(\phi_B) \\
			P_{TN} = & \frac{1}{2}\Bigl(I_2(\theta_B)\sin(2\phi_B) + i\mathit{H_m}(\theta_B)\Bigr) \label{eq:endobservedPs}
\end{align}
\par
We can now go back to the results and look at the amplitude of the fitting parameters $A$, $B$, $C$, and $D$ (Equations \ref{eq:fitsstart} - \ref{eq:fitsend} and Table \ref{table:fitvalues}), as a function of $\theta_B$, shown in Figure \ref{fig:Theta_Residuals}. The error bars are the standard error from the linear least squares fitting. Figure \ref{fig:Theta_Residuals} shows a remarkable amount of variety in the power anisotropy of the different tensor elements with $\theta_B$, including that all of the diagonal terms have less power in the field parallel direction $(\theta_B \rightarrow 0^{\circ})$ than in the perpendicular direction, recovering the results of \cite{Bieber96} and \cite{Horbury08} that the total power is anisotropic. We can see again that there are only 4 measurably distinct non-zero real parameters $A_{RR}$, $A_{TT} = A_{NN}$, $B_{TT} = -B_{NN} = B_{TN}$, and $B_{RT} = B_{RN}$ and two non-zero imaginary parameters $D_{RT} = -D_{RN}$ and $C_{TN}$ and that their dependence on $\theta_B$ is similar. Rather than interpret this in terms of $P^{RTN}(f, \theta_B)$ the direct link with the projections of the four scalar functions is more obvious if we consider the six $\phi_B$-independent functions in Equations \ref{eq:PRRNoPhi}, \ref{eq:TraceNoPhi}, \ref{eq:PRTRNNoPhi}, \ref{eq:PTTNNTNNoPhi}, \ref{eq:I3} and \ref{eq:Hm}. 
\par
\begin{figure*}[htb]%
\centering
\includegraphics[width=\textwidth]{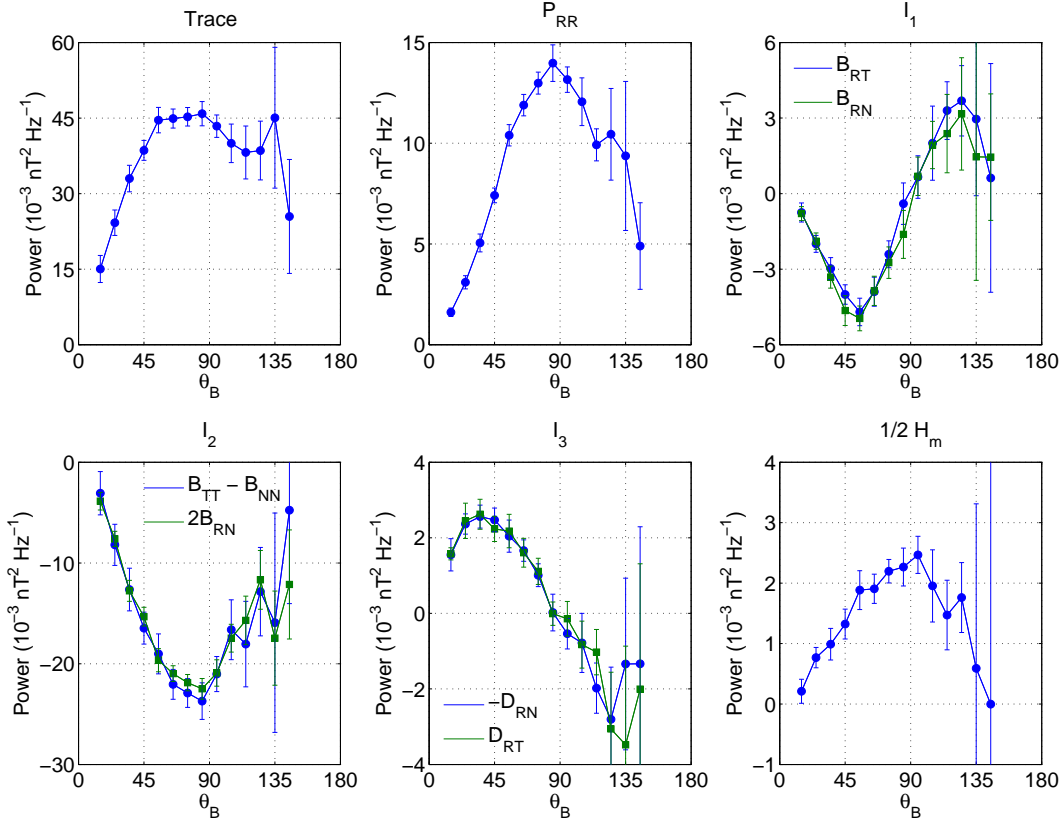}
\caption{Power distributions of the six $\phi_B$ independent integrals as a function of $\theta_B$ at $f = 0.098$ Hz. Error bars are calculated as the error on the fits in the $\phi_B$ direction as in Figure \ref{fig:Theta_Residuals}.}%
\label{fig:I_Functions}%
\end{figure*}

We can extract the six $\phi_B$-independent functions from the fits to the data in Section 2, and can even measure $I_{1-3}$ in two different ways to cross check the results:
\begin{align}
P_{RR}(\theta_B) = & A_{RR}(\theta_B)\label{eq:startObservedAmplitudes}\\
Trace(\theta_B) = & A_{RR}(\theta_B) + A_{TT}(\theta_B) + A_{NN}(\theta_B) \\
H_m(\theta_B) = & 2C_{TN}(\theta_B) \\
I_1(\theta_B) = & B_{RT}(\theta_B) = B_{RN}(\theta_B) \\
I_2(\theta_B) = & B_{TT}(\theta_B) - B_{NN}(\theta_B) = 2 B_{TN}(\theta_B) \\
I_3(\theta_B) = & -D_{RT}(\theta_B) =  D_{RN}(\theta_B) \label{eq:endObservedAmplitudes}
\end{align} 
Figure \ref{fig:I_Functions} shows these six $\phi_B$-independent functions directly measured from the fitting of the power spectral tensor. The two different ways of measuring $I_{1-3}$ all agree with each other remarkably well, which is not required in general, but shows that the turbulence is solenoidal and that the \cite{Oughton97} theory applies and our subsequent derivations are correct. It is also interesting to note that although $I_1$ and $I_3$ appear sinusoidal in $2\theta_B$ upon closer inspection the peaks of the power are shifted from $45^{\circ}$. 
\par
By considering the properties of $Trace$, $P_{RR}$, $I_1$ and $I_2$ we can deduce some important properties of the underlying $\mathit{Tor}(\vct{k})$, $\mathit{Pol}(\vct{k})$ and $\mathit{C}(\vct{k})$ functions. As discussed above we observe that within errors $I_1$ has no imaginary part. We also know that $\mathit{Tor}(\vct{k})$ and $\mathit{Pol}(\vct{k})$ are not both identically zero since both the $Trace$ and $P_{RR}$ are not zero. If we write out the complex terms in Equation \ref{eq:PRTRNNoPhi} for $I_1$ we find that the imaginary part contains terms in $\mathit{Tor}(\vct{k})$, $\mathit{Pol}(\vct{k})$ and $\mathit{C}(\vct{k})$ all with pre-factors that are odd functions of $k_y$. One possible way for $Im[I_1] = 0$ is therefore if $\mathit{Tor}(\vct{k})$ and $\mathit{Pol}(\vct{k})$ are certainly even functions of $k_y$, that is mirror-symmetric about the ($\vct{V},\vct{B}$) plane, since they then integrate to zero. $\mathit{Tor}(\vct{k})$ and $\mathit{Pol}(\vct{k})$ are shown to be even in \cite{Oughton97} and so our results are in accord with theirs. Axi-symmetry about $\hat{\vct{b}}$ is a stronger conclusion not proven, but consistent with mirror-symmetry in $k_y$. Similarly $\textit{C}(\vct{k})$ must be even in $k_y$, or alternatively it can be zero. However, as shown in \cite{Oughton97}, $\textit{C}(\vct{k})$ is necessarily odd, so that we must conclude $\textit{C}(\vct{k}) = 0$.
\par
Applying the same analysis to the imaginary part of $I_2$ we see that the terms multiplying $\mathit{Tor}(\vct{k})$ and $\mathit{Pol}(\vct{k})$ are also odd functions of $k_y$ and so are integrated to zero by Equation \ref{eq:ReducedPower}, however the terms that multiply $\textit{C}(\vct{k})$ are even in $k_y$ and so mirror-symmetry cannot be used to explain the lack of contribution to the power. Thus since $\mathit{Im}[I_1(f, \theta_B)] = \mathit{Im}[I_2(f, \theta_B)] = 0$ we again conclude that $\textit{C}(\vct{k}) = 0$. 
\par
Continuing to $I_3$ we find that the real part, which involves only the magnetic helicity $\mathit{H}(\vct{k})$, has pre-factors that are odd functions of $k_y$, and the imaginary part has even pre-factors in $k_y$. We observe that $\mathit{Re}[I_3] = 0$ since $F \sim 0$, so again following the same line of reasoning we find that $\mathit{H}(\vct{k})$ is also even in $k_y$ and could be axi-symmetric about $\hat{\vct{b}}$.
\par
This is as far as we can conveniently proceed with the analysis in {\it RTN} coordinates. We have used the geometrically induced $\phi_B$ dependence to draw conclusions about the symmetries of $\mathit{Tor}(\vct{k})$, $\mathit{Pol}(\vct{k})$ and $\mathit{H}(\vct{k})$ and to show that $\textit{C}(\vct{k}) = 0$. The {\it RTN} coordinate system, however, also imposes strong geometrical $\theta_B$ dependencies, as can be seen in Figures \ref{fig:Theta_Residuals} and \ref{fig:I_Functions}. The large imposed $\theta_B$ dependence of $P_{RR}$, and $I_{1-3}$ (Equations \ref{eq:PRRNoPhi}, \ref{eq:PRTRNNoPhi}, \ref{eq:PTTNNTNNoPhi} and \ref{eq:I3}) make it hard to extract further information about $\mathit{Tor}(\vct{k})$, $\mathit{Pol}(\vct{k})$ and $\mathit{H}(\vct{k})$ from them. Therefore, at this point we convert the observed power spectral tensor in {\it RTN}, to the $\hat{\vct{b}}$-aligned {\it XYZ} coordinates. 
\par
\begin{figure*}[!htb]%
\centering
\includegraphics[width=\textwidth]{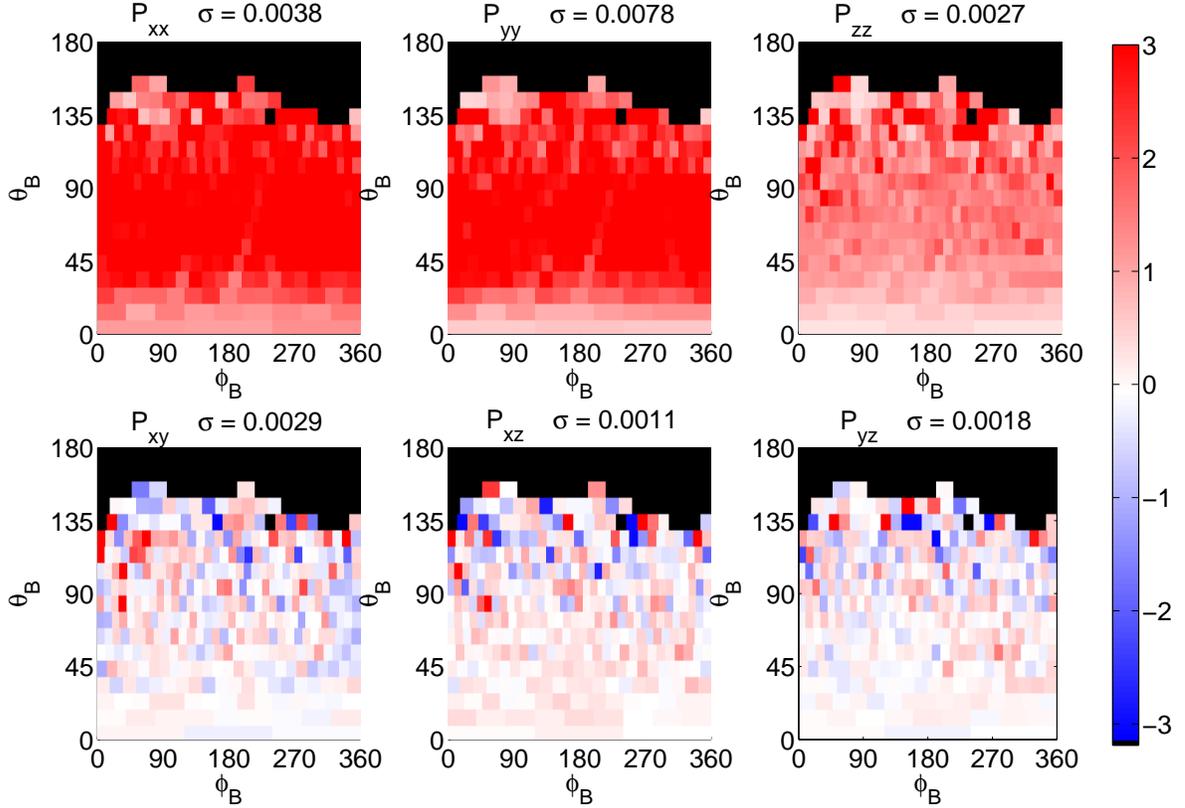}%
\caption{The real component of the power spectral tensor from Ulysses magnetic field data at $f = 0.098$ Hz in field-aligned-coordinates. Black areas represent bins that have fewer than 10 points in them. Red represents positive and blue negative contributions to the power, with white being zero. The color scale has been scaled to the standard deviation of power in each map individually, the value of which is shown above each panel. Compared to Figure \ref{fig:3} in {\it RTN} coordinates much of the variation has been removed.}%
\label{fig:Pxyz2D}%
\end{figure*}
In the field-aligned {\it XYZ} coordinates defined in Equations \ref{eq:ez} - \ref{eq:es}, the $\hat{\vct{t}}$ and $\hat{\vct{p}}$ vectors are:
\begin{eqnarray}
\hat{\vct{t}} &=& -\frac{k_y}{k_{\perp}}\vct{e}_x + \frac{k_x}{k_{\perp}} \vct{e}_y \label{txyz} \\
\hat{\vct{p}} &=& -\frac{k_x k_z}{k k_{\perp}}\vct{e}_x - \frac{k_y k_z}{k k_{\perp}} \vct{e}_y + \frac{k_{\perp}}{k}\vct{e}_z. \label{pxyz}
\end{eqnarray}
importantly they do not involve $\theta_B$ or $\phi_B$ at all. The transformation of a wavelet coefficient in {\it RTN} to {\it XYZ} coordinates is
\begin{equation}
w_i = \sum_j{(\vct{e}_i \cdot \vct{j})w_j} = \sum_j{\vct{M}_{ij}w_j} \label{transform}
\end{equation}
where $j = R$, $T$, $N$, and $i = x$, $y$, $z$ and
\begin{equation}
\vct{M} = \left( \begin{array}{ccc} 
\sin\theta_B & -\cos\theta_B\cos\phi_B & -\cos\theta_B\sin\phi_B \\
0 & \sin\phi_B & -\cos\phi_B \\
\cos\theta_B & \sin\theta_B\cos\phi_B & \sin\theta_B\sin\phi_B 
\end{array} \right)
\label{TransformMatrix}
\end{equation}
We use this matrix to transform the observed reduced power spectral tensor from the {\it RTN} coordinates of the data to the {\it XYZ} coordinates aligned with $\hat{\vct{b}}$ below, using: 
\begin{equation}
\vct{P}^{xyz}(f,\theta_B) = \vct{M} \vct{P}^{RTN}(f,\hat{\vct{b}}) \vct{M}^T. \label{eq:PSTtransform}
\end{equation}
In field-aligned coordinates the transformed power spectral tensor should have no $\phi_B$ dependence and simpler $\theta_B$ dependence than shown in Figures \ref{fig:3}-\ref{fig:Phi_Fits} since the $\hat{\vct{t}}$ and $\hat{\vct{p}}$ vectors in {\it XYZ} are no longer dependent on these angles.
\par
The expressions for $P^{xyz}(\vct{k})$ using the definitions of $\hat{\vct{t}}$ and $\hat{\vct{p}}$ in Equations \ref{txyz} and \ref{pxyz} are:
\begin{align}
P_{xx}(\vct{k}) = &\  \frac{k_y^2}{k_{\perp}^2}\mathit{Tor}(\vct{k}) + \frac{k_x^2k_z^2}{k_{\perp}^2k^2}\mathit{Pol}(\vct{k}) + 2\frac{k_xk_yk_z}{k_{\perp}^2k}\mathit{C}(\vct{k}) \label{eq:Pxx}\\
P_{yy}(\vct{k}) = &\  \frac{k_x^2}{k_{\perp}^2}\mathit{Tor}(\vct{k}) + \frac{k_y^2k_z^2}{k_{\perp}^2k^2}\mathit{Pol}(\vct{k}) - 2\frac{k_xk_yk_z}{k_{\perp}^2k}\mathit{C}(\vct{k}) \label{eq:Pyy}\\
P_{zz}(\vct{k}) = &\  \frac{k_{\perp}^2}{k^2}\mathit{Pol}(\vct{k}) \label{eq:Pzz}\\
P_{xy}(\vct{k}) = &\  \frac{k_xk_y}{k_{\perp}^2}\mathit{Tor}(\vct{k}) + \frac{k_xk_yk_z^2}{k_{\perp}^2k^2}\mathit{Pol}(\vct{k}) + \frac{(k_y^2 - k_x^2)k_z}{k_{\perp}^2k}\mathit{C}(\vct{k}) + \mathit{i}k_z\mathit{H}(\vct{k}) \label{eq:Pxy}\\
P_{xz}(\vct{k}) = &-\frac{k_xk_z}{k^2}\mathit{Pol}(\vct{k}) - \frac{k_y}{k}\mathit{C}(\vct{k}) - \mathit{i}k_y\mathit{H}(\vct{k}) \label{eq:Pxz}\\
P_{yz}(\vct{k}) = &-\frac{k_yk_z}{k^2}\mathit{Pol}(\vct{k}) + \frac{k_x}{k}\mathit{C}(\vct{k}) + \mathit{i}k_x\mathit{H}(\vct{k}) \label{eq:Pyz} 
\end{align}
Although there is explicit dependence on components of $\vct{k}$ in every term there are no $\theta_B$ or $\phi_B$ dependent geometrical pre-factors. $\textit{C}(\vct{k})$ and terms odd in $k_y$ are shown for completeness, although we know from the {\it RTN} analysis that they do not contribute to $P^{xyz}(f, \theta_B)$ in the solar wind. The reduction to $P(f, \theta_B)$ using Equation \ref{eq:ReducedPower} introduces dependence on $\theta_B$ if there is any anisotropy in the scalar functions.
\par
Figure \ref{fig:Pxyz2D} shows the same results as Figure \ref{fig:3}, the real part of the power spectral tensor, converted to {\it XYZ} coordinates. There is no $\phi_B$ dependence in any of the elements and no sinusoidal-like dependence on $\theta_B$. This result, combined with the strong $\phi_B$ and $\theta_B$ dependence of $P^{RTN}$ implies that we have correctly identified the direction of $\hat{\vct{b}}$ using the local wavelet averaging method. If we had identified $\hat{\vct{b}}$ incorrectly with a systematic error then there would be a sinusoidal dependence on $\theta_B$ in $P^{xyz}$ and if there was a random error smoothing out the variations we could not have measured the precisely predicted $\phi_B$ dependence of $P^{RTN}$. 
\par
\begin{figure*}[!htb]%
\centering
\includegraphics[width=\textwidth]{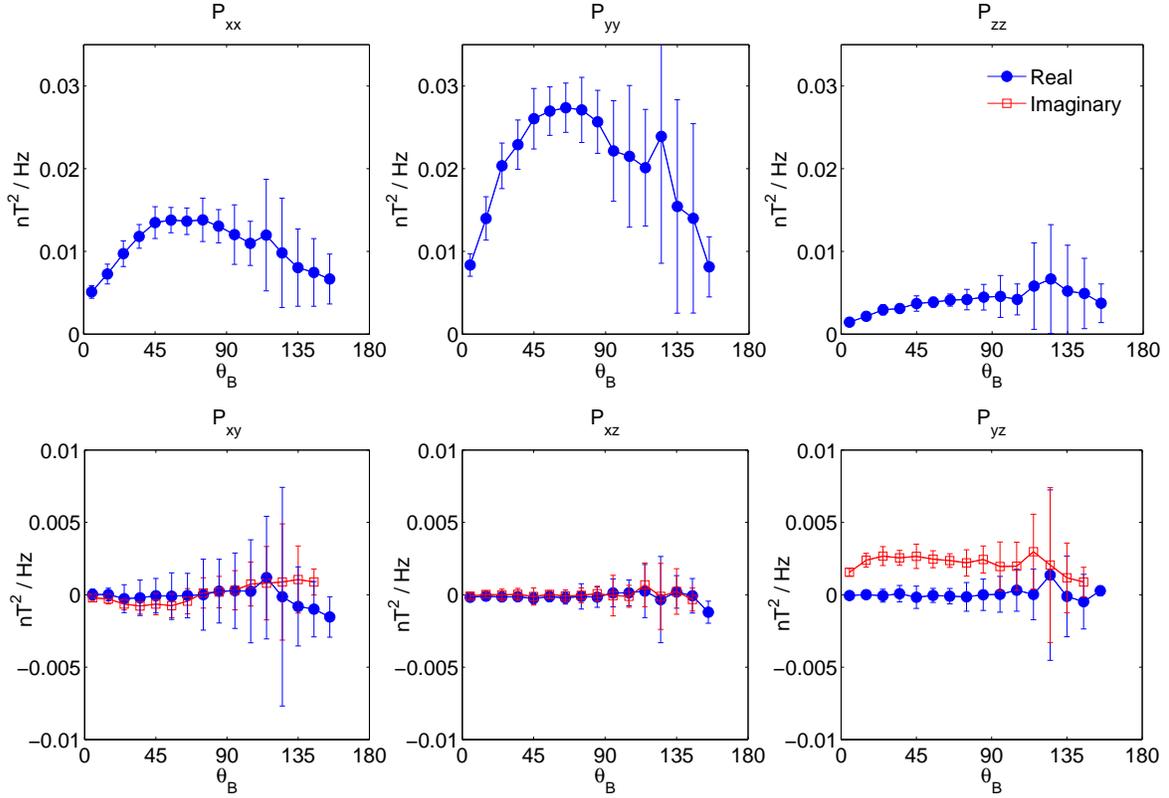}%
\caption{Power anisotropy of each power spectral tensor element as a function of $\theta_B$ in magnetic-field-aligned coordinates at $f = 0.098$ Hz. The error bars are calculated as the error on the mean in each $\theta_B$ bin which have no dependence on $\phi_B$.}%
\label{fig:PowerAnisoXYZ}%
\end{figure*}

Figure \ref{fig:PowerAnisoXYZ} shows the $\theta_B$ dependence of each independent element in both the real and imaginary parts; the error bars are calculated as the error on the mean of all data contributing to each bin in $\theta_B$. This figure allows us to make further deductions about the scalar functions using a similar analysis procedure to that of the {\it RTN} tensor previously. First, the real parts of the off-diagonal elements are all observed to be within errors of zero. Looking at Equations \ref{eq:Pxy} - \ref{eq:Pyz} we see that $\mathit{Tor}(\vct{k})$ only appears combined with an odd function of $k_y$ and so we have rediscovered its mirror-symmetry, $\mathit{Pol}(\vct{k})$ is also combined with an odd function of $k_y$ in two of the elements but $P_{xz}(\vct{k}) \propto k_xk_z\mathit{Pol}(\vct{k})$ so $\mathit{Pol}(\vct{k})$ must be predominantly 2D, that is mostly confined to $|k_z| \sim 0$. Finally $\mathit{C}(\vct{k})$ is multiplied by $k_y$ and $k_y^2$ in $P_{xz}(\vct{k})$ and $P_{xy}(\vct{k})$ respectively and so must be zero, as previously discovered.
\par
The imaginary parts $\mathit{Im}[P_{xy}]$ and $\mathit{Im}[P_{xz}]$ are within errors of zero. $\mathit{Im}[P_{xz}]$ is combined with an odd function of $k_y$ so again the data imply that $\mathit{H}(\vct{k})$ is mirror-symmetric in $k_y$. $\mathit{Im}[P_{xy}] = 0$ implies a further symmetry of $\mathit{H}(\vct{k})$ in $k_z$, similar to $\mathit{Pol}(\vct{k})$ this implies that in the inertial range $\mathit{H}(\vct{k})$ is mostly associated with $\vct{k}$s in the plane perpendicular to $\vct{B}$. At the highest frequencies studied here $\mathit{Im}[P_{xy}]$ becomes anisotropic (a very small signature may be visible in Figure \ref{fig:PowerAnisoXYZ} although it is indistinguishable from 0 when errors are considered and is zero in the inertial range in general), however this is probably associated with plasma instabilities near the ion gyroscale \citep{He2011, Podesta2011}. Only $\mathit{Im}[P_{yz}] \propto k_x\mathit{H}(\vct{k})$ has finite value in general, although it is much smaller than the Trace and has much weaker $\theta_B$ dependence than $\mathit{Im}[P_{TN}]$.
\par
Moving on to the diagonal elements, we see that as in {\it RTN} coordinates the diagonal elements are anisotropic with more power at $\theta_B \sim 90^{\circ}$ than at $\theta_B \sim 0^{\circ}$. They are ordered in power with $P_{yy} > P_{xx} > P_{zz}$ seeming to show a 3D anisotropy with the most power perpendicular to the ($\vct{V},\vct{B}$) plane, then intermediate power perpendicular to $\vct{B}$ but in the ($\vct{V},\vct{B}$) plane and finally the least power parallel to $\vct{B}$, exactly as first shown by \cite{Belcher71} and in agreement with the well established results that the solar wind is anisotropic with $0 < P_{zz} < P_{xx} + P_{yy}$ \citep[e.g.][]{Matthaeus90, Dasso05, Matthaeus05, Osman07}. Looking at Equations \ref{eq:Pxx}-\ref{eq:Pzz} we see that this ordering implies that $0 < \frac{k_{\perp}^2}{k^2}\mathit{Pol}(\vct{k}) < \mathit{Tor}(\vct{k}) + \frac{k_{||}^2}{k^2}\mathit{Pol}(\vct{k})$. One way this can be achieved is if $\mathit{Pol}(\vct{k}) < \mathit{Tor}(\vct{k})$, although this is not required.
\par
In the assumption that the turbulence is 2D the reduced power $P_{yy}(f, \theta_B) > P_{xx}(f, \theta_B)$ is due to the reduction integral (Equation \ref{eq:ReducedPower}) \citep{Bieber96, Turner11}. We can now make a stronger statement than in the previous paragraph, we have shown that $\mathit{Pol}(\vct{k})$ is mostly due to approximately 2D wavevectors and so does not contribute strongly to either of these terms since they both contain $k_z^2\mathit{Pol}(\vct{k})$, thus the observed power is from the purely Alfv\'{e}nic $\mathit{Tor}(\vct{k})$ fluctuations. $P_{xx}(f, \theta_B)$ and $P_{yy}(f, \theta_B)$ are reduced in Equation \ref{eq:ReducedPower} as $\propto k_y^2\mathit{Tor}(\vct{k})$ and $\propto k_x^2\mathit{Tor}(\vct{k})$ respectively and it is interesting to note that $P_{yy}(f, \theta_B) / P_{xx}(f, \theta_B)$ is approximately $2$ at all values of $\theta_B$ at this frequency. As \cite{Turner11} showed the reduction integral applied to a power spectrum of 2D fluctuations results in a constant factor proportional to the spectral index of the turbulence, so this apparent anisotropy is a feature associated with sampling along a single cut through the data and the results of \cite{Belcher71, Bieber96, Turner11} now have unified explanation. We have also separated the reduced form of the Alfv\'{e}nic $\mathit{Tor}(\vct{k})$ with two different projections from the reduced pseudo-Alfv\'{e}nic $\mathit{Pol}(\vct{k})$ in the observations. They have different power levels and anisotropy since $P_{zz}(f, \theta_B)$ is very different in magnitude and shape from $P_{yy}(f, \theta_B)$, which is approximately $2P_{xx}(f, \theta_B)$.

\section{Summary and Conclusions}

We have observed all nine elements of the reduced power spectral tensor of MHD-scale fluctuations in fast solar wind using wavelet transforms of magnetic field observations by the Ulysses spacecraft. Each element of the tensor is resolved using angle coordinates $\theta_B$ and $\phi_B$ at a single frequency $f = 0.098$ Hz. The signal is anisotropic and depends on the direction of the local mean magnetic field. This anisotropy can be seen in Figures \ref{fig:3} and \ref{fig:4} and it is quantified in the $\phi_B$ direction by fitting sinusoidal functions in Figure \ref{fig:Phi_Fits} with the amplitudes given in Table \ref{table:fitvalues}. These show that within errors we observe only six $\phi_B$-independent power amplitudes in the solar wind.
\par
We explain the generation of this anisotropy analytically by applying a scalar field and tensor description of solenoidal turbulence \citep{Oughton97}. We choose our scalar functions so that they represent the toroidal ($\mathit{Tor}(\vct{k})$) and poloidal ($\mathit{Pol}(\vct{k})$) fluctuations with respect to the local mean magnetic field direction $\hat{\vct b}$, and their in- and out-of-phase correlations ($\mathit{C}(\vct{k})$ and $\mathit{H}(\vct{k})$). We then convert this representation into the spacecraft data coordinate system where the toroidal $\hat{\vct{t}}$ and poloidal $\hat{\vct{p}}$ directions are expressed in terms of heliocentric {\it RTN} coordinates (Equations \ref{eq:tarray} and \ref{eq:parray}). Applying the reduction integral to the four scalar fields in conjunction with the appropriate dyadics of $\hat{\vct{t}}$ and $\hat{\vct{p}}$ we derive the dependence of the reduced power tensor $P_{ij}(f, \hat{\vct{b}})$ on the four scalar fields. 
\par
While we do not know the analytical form of any of the four scalar fields, the geometrical dependence on $\phi_B$ in the {\it RTN} coordinate system is independent of the reduction integral and simple sinusoidal dependences on $\phi_B$ are found. To simplify the equations we gathered terms with no, first harmonic or second harmonic dependence on $\phi_B$. We found six combinations of the reduced power spectral tensor that analytically have precise sinusoidal $\phi_B$ dependence, matching up with the six observed independent power amplitudes. These are the $Trace$, $P_{RR}$ and $\mathit{Im}[P_{TN}]$ elements of the reduced power spectral tensor which do not depend on $\phi_B$, and the $I_1$, $I_2$,and $I_3$ amplitudes of the harmonic components in the other elements, defined in section 4. Observations of these from the solar wind do indeed have no $\phi_B$ dependence confirming the derivations and their $\theta_B$ dependence is shown in Figure \ref{fig:I_Functions}. From these results we draw several conclusions:
\begin{enumerate}

	\item The $\phi_B$ dependence of the data (Figure \ref{fig:Phi_Fits}) follows the form of the power spectral tensor derived for solenoidal fluctuations transformed from field aligned coordinates in to {\it RTN} coordinates.
	
	\item $\mathit{Im}[I_1] = \mathit{Im}[I_2] = 0$ implies that the turbulent power in $\mathit{Tor}(\vct{k})$ and $\mathit{Pol}(\vct{k})$ is an even function of $k_y$ and so is mirror-symmetric about the ($\vct{V},\vct{B}$) plane, and thus it is likely to be axi-symmetric about $\hat{\vct{b}}$.
	
	\item The integrand of $\mathit{Im}[I_1] \propto k_y\mathit{C}$ and the integrand of $\mathit{Im}[I_2] \propto k_y^2\mathit{C}$ and both are observed to be zero when integrated; thus the scalar function $\mathit{C}(\vct{k}) = 0$.
	
\end{enumerate}
\par
The strong $\theta_B$ dependence observed in all panels of Figure \ref{fig:I_Functions} arises from the integrals over a combination of any actual anisotropy of the turbulence \citep{Bieber96, Dasso05, GS95, GS97, Matthaeus96, Matthaeus98, Oughton98, Oughton11} and geometrical effects. This can be seen in Equations \ref{eq:PRRNoPhi}-\ref{eq:PTTNNTNNoPhi} and \ref{eq:I3} as their dependence on the scalar fields, the coordinate transformed unit vectors $\hat{\vct{t}}$ and $\hat{\vct{p}}$, and the complex functions $Z_t$ and $Z_p$ (Equations \ref{eq:Zt}, \ref{eq:Zp}). Since we cannot make simplifying assumptions such as symmetries in the $\theta_B$ direction, and since we do not know the analytical form of the scalar functions, further progress in the {\it RTN} coordinate system is difficult. 
\par
We therefore transform the observed $P(f, \hat{\vct{b}})$ into magnetic-field-aligned coordinates {\it XYZ}. All the $\phi_B$ variation in power disappears in agreement with the theoretical prediction (Equations \ref{eq:Pxx}-\ref{eq:Pyz}). The observations have completely reproduced the theoretical prediction for the $\phi_B$ dependence in {\it RTN} and the independence in {\it XYZ} coordinates, of the power, confirming that our measurement of $\hat{\vct{b}}$ using the local mean magnetic field is an axis of symmetry for the ensemble average. This is a strong justification for using the local mean field when studying anisotropy in turbulence since our results indicate that this direction has a strong influence on the symmetry of the scalar functions.  
\par
By comparing the analytically derived field-aligned power spectral tensor elements (Equations \ref{eq:Pxx}-\ref{eq:Pyz}) with the data in Figures \ref{fig:Pxyz2D} and \ref{fig:PowerAnisoXYZ} we can draw further conclusions:
\begin{enumerate}[resume]

	\item The Alfv\'{e}nic $\mathit{Tor}(\vct{k})$ fluctuations can be separated from the pseudo-Alfv\'{e}nic $\mathit{Pol}(\vct{k})$ fluctuations since $P_{zz}(f, \theta_B)$ is a function of $\mathit{Pol}(\vct{k})$ alone. 
	
	\item $\mathit{Pol}(\vct{k})$ is measurable and has a different power anisotropy with respect to $\theta_B$ than $\mathit{Tor}(\vct{k})$. Since observationally $P_{zz} < P_{xx}, P_{yy}$ it also seems likely that $\mathit{Pol}(\vct{k}) < \mathit{Tor}(\vct{k})$.
	
	\item $\mathit{Pol}(\vct{k})$ is even in $k_y$ and therefore it is likely to be axi-symmetric about $\hat{\vct b}$ since the observed reduced real parts of $P_{xy}(f, \theta_B)$ and $P_{yz}(f, \theta_B)$ are zero and $P_{xy}(\vct{k}) \propto P_{yz}(\vct{k}) \propto k_y\mathit{Pol}(\vct{k})$.
	
	\item $\mathit{Pol}(\vct{k})$ is mostly due to fluctuations with $|k_z| \sim 0$ because the observed reduced $P_{xz}(f, \theta_B)$ is zero and $P_{xz}(\vct{k}) \propto k_z\mathit{Pol}(\vct{k})$.
	
	\item	$\mathit{C}(\vct{k}) = 0$ is confirmed by all real off-diagonal elements being zero since combined they have both odd ($P_{xz}$ and $P_{yz}$) and even ($P_{xy}$) pre-factors in $k_y$ and $k_x$.
	
	\item $\mathit{H}(\vct{k})$ is even in $k_y$ and therefore likely axi-symmetric about $\hat{\vct b}$, since the imaginary off-diagonal element $P_{xz}(f, \theta_B)$ is zero and $P_{xz}(\vct{k}) \propto k_y$.

	\item $\mathit{H}(\vct{k})$ comes from fluctuations that have wave vectors in the plane perpendicular to $\vct{B}$ with power confined around $|k_z| \sim 0$ since the imaginary part of the off-diagonal element $P_{xy}$ is zero and $P_{xy}(\vct{k}) \propto k_z$

\end{enumerate}
\par
A physical interpretation of these results is that turbulence in the solar wind is made up of mostly toroidal fluctuations that are anisotropic. The observed $P_{zz}$ and therefore $\mathit{Pol}(\vct{k})$ is compatible with solenoidal fluctuations as in Equation \ref{eq:Pzz}, however, a spectrum of $|\vct{B}|$ fluctuations is observed in the fast solar wind, so we cannot rule out compressible plasma fluctuations as a source of this variation. If we consider the results in terms of a superposition of polarized fluctuations then $\mathit{Tor}(\vct{k}) > \mathit{Pol}(\vct{k})$ implies the fluctuations must be elliptical on average \citep[e.g.][]{Chen11, Mallet11}. $\mathit{H}(\vct{k}) \neq 0$, implied by the finite values of $P_{TN}(f, \theta_B)$ and $P_{yz}(f, \theta_B)$, means that the $\mathit{Tor}(\vct{k})$ and $\mathit{Pol}(\vct{k})$ fluctuations are partially correlated and there is a polarization ellipse. $\mathit{C}(\vct{k}) = 0$ implies that the ensemble averaged polarization ellipse is oriented along $\hat{\vct{t}}$ or $\hat{\vct{p}}$ \citep{Chandrasekhar60} but since the solar wind is not entirely coherent waves this must be a result of a superposition of polarization ellipse orientations that average to zero. Thus the ensemble average turbulence is similar to partially polarized, partly natural (incoherent) light \citep{Chandrasekhar60}.
\par
Recently \cite{Turner11} showed that the difference in power $P_{yy}(f) > P_{xx}(f)$ can arise from the reduction of an axi-symmetric 2D turbulence. Here we have shown why they find agreement between a superposition of 2D Alfv\'{e}n waves, numerical MHD simulations and the solar wind: the Alfv\'{e}nic $\mathit{Tor}(\vct{k})$ dominates the pseudo-Alfv\'{e}nic $\mathit{Pol}(\vct{k})$ contribution to both $P_{xx}$ and $P_{yy}$ when they are reduced and so observations of these terms appear Alfv\'{e}nic, even if pseudo-Alfv\'{e}nic fluctuations exist. The results we have shown here set the work of \cite{Turner11} in the wider physical context of the full turbulent power spectral tensor.
\par
The observation process demonstrated in this paper can be repeated at many different scales rather than just one so the scaling of the $\phi_B$ invariant functions and the field aligned power spectral tensor can be measured. This may help test different theories for anisotropic turbulence if theoretical predictions for the scaling of the scalar functions are made and we intend to present such an analysis in the near future. Finally, this work also demonstrates that care should be taken when using off-diagonal terms from the power spectral tensor to observe physical phenomena. For example, in work such as \cite{He2011} and \cite{Podesta2011}, the magnetic helicity is measured as $\mathit{Im}[P_{TN}]/$Trace, however the trace has its own power anisotropy \citep{Horbury08, Podesta09, Luo10, Wicks10} which depends mostly on $\mathit{Tor}(\vct{k})$ which we have shown is different from the anisotropy of $\mathit{H}(\vct{k})$ alone. Thus dividing by the trace introduces or removes apparent anisotropy from these results. Furthermore from the work presented here we can see that $P_{xy}$, $P_{xz}$ and $P_{yz}$ contain different projections of $\mathit{H}(\vct{k})$, from which we may learn more about the properties and symmetries of the helicity induced by solenoidal turbulence and instabilities.

\label{lastpage}

%
%


\end{document}